\documentclass[runningheads,fleqn]{cl2emult}

\usepackage{makeidx}  
\usepackage{graphicx} 
\usepackage{subeqnar} 
\usepackage{multicol} 
\usepackage{cropmark} 
\usepackage{phys}     
\makeindex            

\newcommand\la{\langle}
\newcommand\ra{\rangle}
\newcommand\scro{{\cal O}}
\newcommand{\beq}{\begin{equation}}
\newcommand{\eeq}{\end{equation}}
\newcommand{\be}{\begin{equation}}
\newcommand{\ee}{\end{equation}}
\newcommand{\beqn}{\begin{eqnarray}}
\newcommand{\eeqn}{\end{eqnarray}} 
\def\reff#1{(\ref{#1})}

\begin{document}
\title*{Monte Carlo Simulation of the Three-dimensional Ising Spin Glass}
\toctitle{Monte Carlo Simulation of the Three-dimensional\protect\newline
 Ising Spin Glass}
%
%
\titlerunning{Monte Carlo Simulation of the 3D Ising Spin Glass}
%
\author{\underline{Matteo Palassini}\inst{1}
\and Sergio Caracciolo\inst{2}}
\authorrunning{M.~Palassini and S.~Caracciolo}
%
%
\institute{Department of Physics, University of California, Santa Cruz, California 95064
\and Scuola Normale Superiore and INFN, 56100 Pisa, Italy}

\maketitle              

\vspace{-0.2cm}
\begin{abstract}
We study the 3D Edwards--Anderson model with binary interactions by 
Monte Carlo simulations. Direct evidence of finite--size scaling is
provided, and the universal finite--size scaling functions are 
determined. Using an iterative extrapolation procedure, 
Monte Carlo data are extrapolated to infinite volume 
up to correlation length $\xi \approx 140$. 
The infinite volume data are consistent with both a  
continuous phase transition at finite temperature and 
an essential singularity at finite temperature. 
An essential singularity at zero temperature is excluded.
\end{abstract}

\vspace{-0.6cm}

\section{Introduction} 

Understanding the thermodynamics of spin glasses 
represents a challenging unsolved problem for statistical 
and computational physics \cite{BY}. 
The very exi\-stence of a phase transition in the 
3D Ising spin glass is still unclear.
Previous Monte Carlo (MC) simulations \cite{ogielsky,on3D,KY,maparu,janke} 
give a certain evidence of a  $T_c\neq 0$ 
continuous transition with power-law divergence of $\xi$ at $T\to T_c^+$.
However, they cannot exclude completely 
neither an exponential divergence at $T \to 0^+$ \cite{ogielsky,on3D,KY}, 
nor a line of critical points at $T \le T_c \neq 0$ \cite{KY,janke,iniguez}, 
which implies an exponential divergence at $T\to T_c^+$.
Furthermore, the important topic of 
finite-size scaling (FSS) has not been
much investigated for spin glasses.

Here, we study the $3D$ Ising spin glass by 
MC simulations of moderate--size systems in the 
paramagnetic phase (less hampered by long equilibration times),
using the powerful FSS method of \cite{fss_greedy}.
In Sect.~2 details on the simulations are given.
In Sect.~3 a direct evidence of FSS, independent
of the nature of the divergence of $\xi$, is provided,
and the {\em universal}\, FSS functions are determined for the first time 
\cite{sg3d}. The MC data are extrapolated to infinite volume up
to $\xi$ one order of magnitude larger than before.  
In {\mbox Sect.~4} the critical behavior of 
the extrapolated data is analyzed.

\section{Model and simulation } 
We consider the $3D$ Edwards--Anderson model, whose Hamiltonian is
\begin{equation}
{\cal H}=- \sum_{\la xy \ra}\sigma_x J_{xy}\sigma_y  \label{model}
\end{equation}
where $\sigma_x$ are Ising spins on a simple cubic lattice of linear size $L$ 
with periodic \newpage \noindent 
boundaries, and $J_{xy}$ are independent random interactions
taking the values $\pm 1$ with equal probabi\-lity.
The sum runs over pairs of nearest neighbor sites.

We simulate with the heath--bath algorithm 
many samples of the model \reff{model} with different ${J_{xy}}$.
From  two independent replicas $(\sigma,\tau)$  with 
the same ${J_{xy}}$, we measure the overlap 
$q_x = \sigma_x \tau_x$   and $q = L^{-3}\sum_x q_x$. 
We compute the {\em second-moment} correlation length using the
following definition:
\beq
  \xi(T,L)   =    
\left( {\, S(0)/S(p) \,-\, 1 \, \over  4 \, \sin^2(p/2) }\right)^{1/2} 
 \label{corr_len_2mom}
\eeq
where $S(k)$ is the Fourier transform of the overlap correlation function
\beq
S(k) = \sum_r e^{ik\cdot r} \la q_x q_{x+r} \ra \;,
\eeq
(the arguments $T,L$ are omitted) and $p = (2\pi/L,0,0)$ is the smallest 
non--zero wave vector. We also measure the spin--glass susceptibility 
$\chi_{SG}(T,L) \equiv L^3 \la q^2 \ra = S(0)$.
The runs are done on a Cray T3E  with a fast code 
(see \cite{sg3d} for details) that exploits the high parallelism 
of spin glass simulations, combining multi--processor 
parallelism with an efficient multi--spin coding technique.
Average speed on a single processor (DEC Alpha EV5, 600 MHz) 
is  $4.5\times 10^7$ spin updates per second, and we typically used 
32 to 128 PEs. We simulated 104 pairs $(T,L)$, $(T,2L)$
with $L$ between  $4$ and $48$. 

\section{Finite size scaling analysis }

\begin{figure}[h]
        \mbox{
	\includegraphics[width=.58\textwidth]{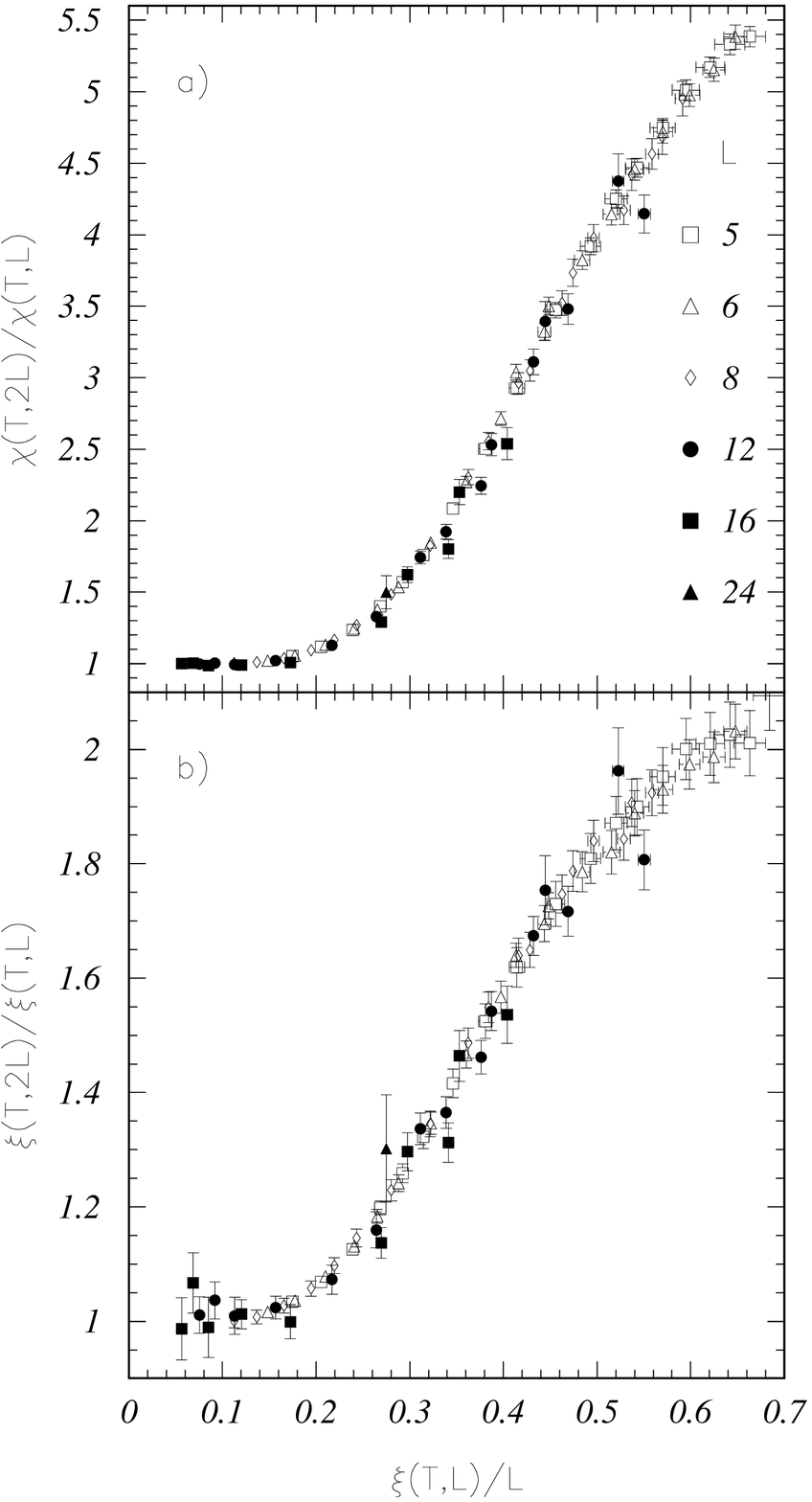}
        }
\hspace{-1.2cm}
        \mbox{
	\includegraphics[width=.58\textwidth]{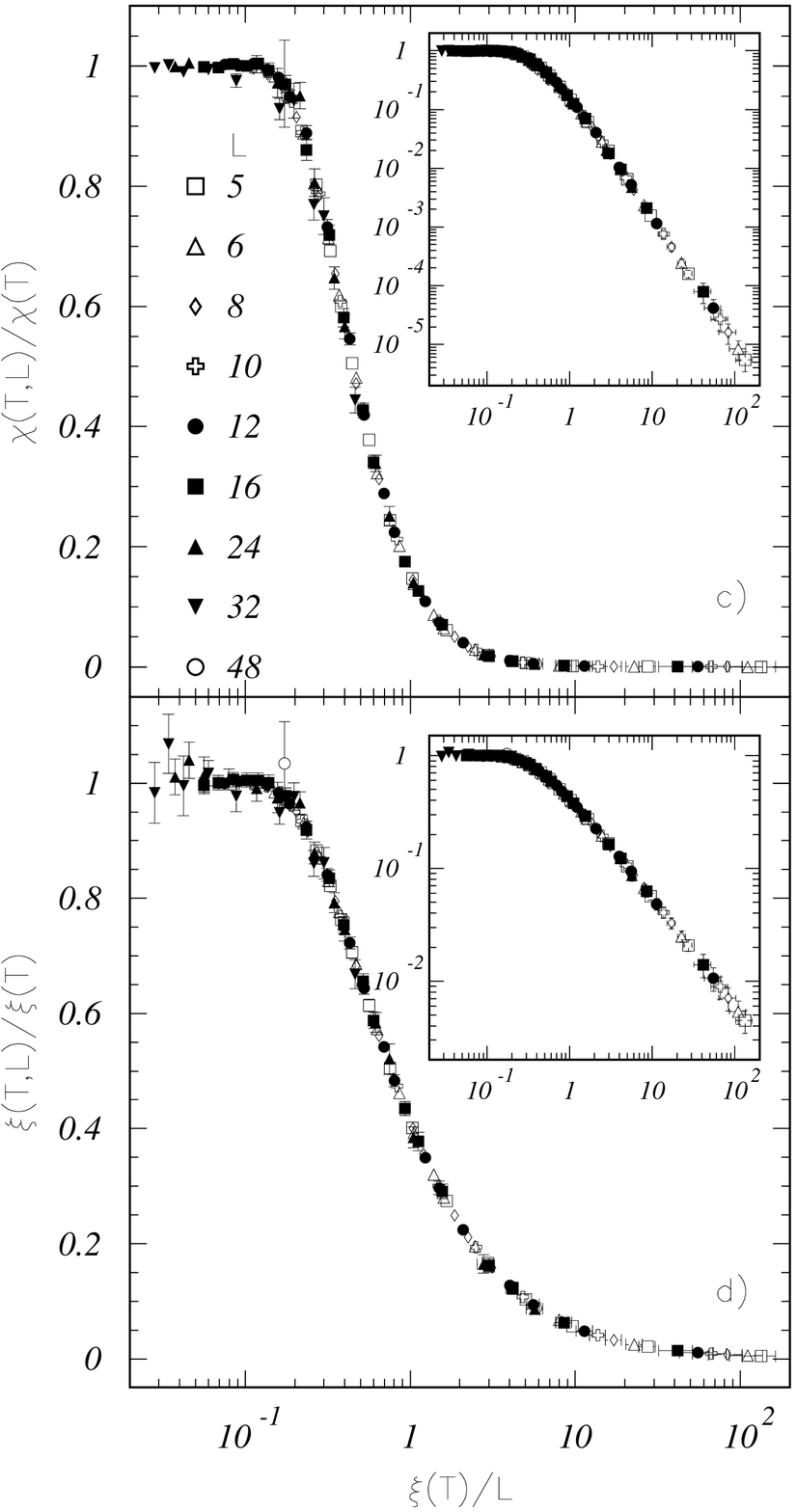}
        }
\hspace{-1.5cm}
\caption{({\bf a,b}) FSS plot with the form in (\protect\ref{eq2}) 
for ({\bf a}) $\scro=\chi_{SG}$ and ({\bf b}) $\scro=\xi$. Error bars  
(estimated with a jackknife procedure) are one standard deviation.
({\bf c,d}) FSS plot with the form in (\protect\ref{eq1}) 
for ({\bf c}) $\scro=\chi_{SG}$ and ({\bf d}) 
$\scro=\xi$. See \protect\cite{fss_greedy} for how to estimate error bars
of the extrapolated data.  }
\label{fssmethod}
\end{figure}

\begin{figure}[h]
        \mbox{
	\includegraphics[width=.48\textwidth]{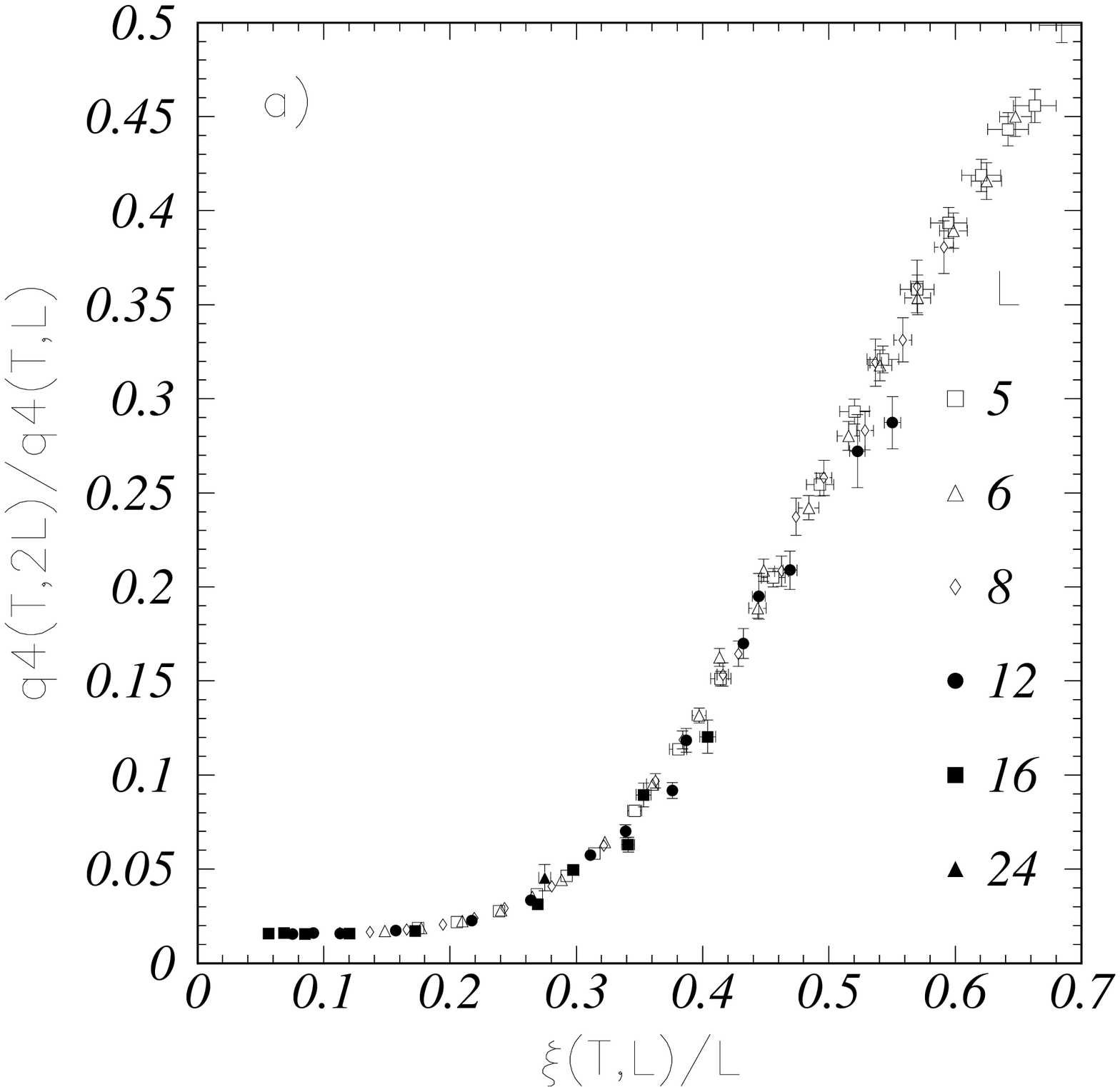}
        }
        \mbox{
	\includegraphics[width=.48\textwidth]{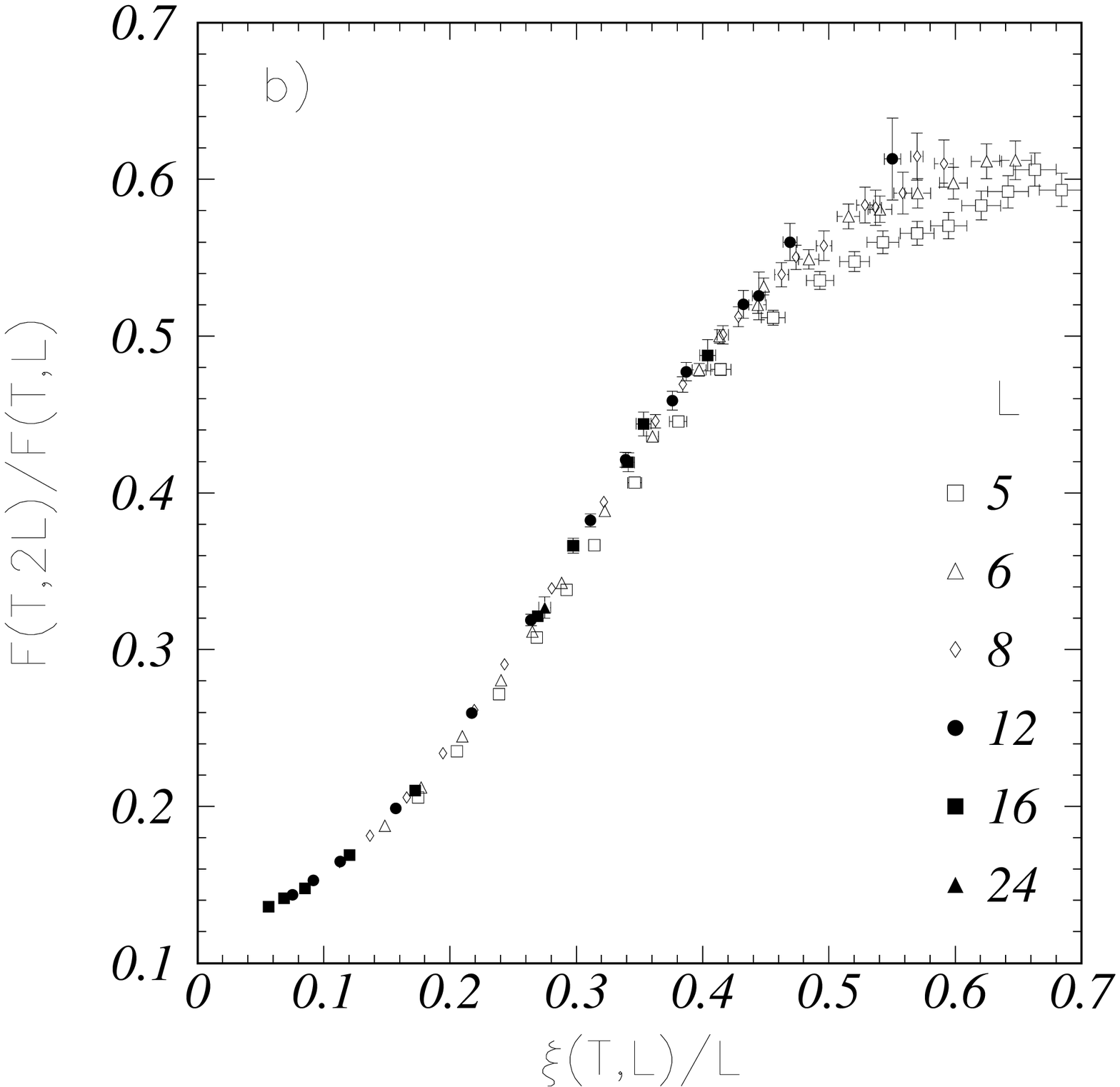}	
        }
\caption{FSS plot with the form in (\protect\ref{eq2}) 
for ({\bf a}) $\scro= q_4 \equiv \la q^4 \ra$ and 
({\bf b}) $\scro=F\equiv V^{-3} S(p)$,
$|p|=2\pi/L$.}
\label{q4}
\end{figure}

According to the FSS hypothesis \cite{Barber_FSS_review}, if
$\scro(T,L)$ is some long-distance observable (for example, 
$\xi(T,L)$ or $\chi_{SG}(T,L)$), then 
\be
   {\scro(T,L) \over \scro(T,\infty)}   \;=\;
   f_{\scro} \Bigl( \xi(T,\infty)/L \Bigr) \, 
   \;
 \label{eq1}
\ee
where $f_{\scro}$ is a universal function and corrections
to FSS are neglected. From \reff{eq1} we obtain a useful relation
involving  only finite--volume observables 
\beq
   {\scro(T,2L) \over \scro(T,L)}   \;=\;
   F_{\scro} \Bigl( \xi(T,L)/L \, \Bigr) \, 
   \;
 \label{eq2}
\eeq
where $F_\scro$ is another universal function.
As shown in Fig.\ref{fssmethod}a,b  and Fig.\ref{q4}, our data
for the observables $\chi_{SG}$, $\xi$, $q_4 \equiv \la q^4 \ra$
and $F \equiv V^{-3} S(p)$ verify the ansatz \reff{eq2}. 
This provides a
direct test of the FSS hypothesis.
Small systematic deviations,
due to corrections to FSS, appear in Fig.\ref{q4}b
for $L=5$ and $\xi(T,L)/L \equiv x > 0.4$. 
We emphasize that FSS is not assumed {\em a priori} and that 
no adjustable parameters are contained in \reff{eq2}.
Furthermore, no particular dependence of the 
observables on the temperature is assumed.

We fit the data in Fig.\ref{fssmethod}a,b to two suitable
functions $F_{\chi_{SG}}$, $F_\xi$ and then
extrapolate the pair $(\xi,\chi_{SG})$  from
$L \to 2L \to 2^2 L \to \ldots \to \infty$  using  \reff{eq2} iteratively.
For any temperature, we verify that 
extrapolations from different $L$ agree within the error bars.
In this procedure, we assumed 
implicitly that \reff{eq2} with a given function $F_\scro$ 
will continue to hold as $L\to\infty$. This assumption could fail if there
is a crossover at large $L$. However, at high $T$  extrapolations 
from small $L$  are consistent with data from large $L$, 
which have almost no finite-size effects, and thus
a crossover is unlikely. We have a good control on the extrapolated 
data up to $\xi \approx 140$; 
at lower temperatures the statistical errors become
quite large, and the data are more sensitive to FSS corrections.

In Fig.\ref{fssmethod}c,d we show that with our extrapolated data 
\reff{eq1} is satisfied remarkably well, providing a further test 
of the method.
If $\scro \sim \xi^{\gamma_\scro/\nu}$ as $\xi \to \infty$, 
then $f_\scro(x)$ in \reff{eq1} must satisfy 
$f_\scro(x)\sim x^{-\gamma_\scro/\nu}$ as $x\to \infty$. 
As shown in the insets of Fig.\ref{fssmethod} (c,d), 
our curves indeed have a power-law asymptotic decay.
We emphasize that all the scaling functions in 
Figs.\ref{fssmethod} and \ref{q4} are {\em universal}.

\section{Nature of the phase transition }

We now compare our extrapolated data with the three different
scenarios compatible with previous simulations.

\begin{figure}[h]
        \mbox{
	\includegraphics[width=.32\textwidth,angle=-90]{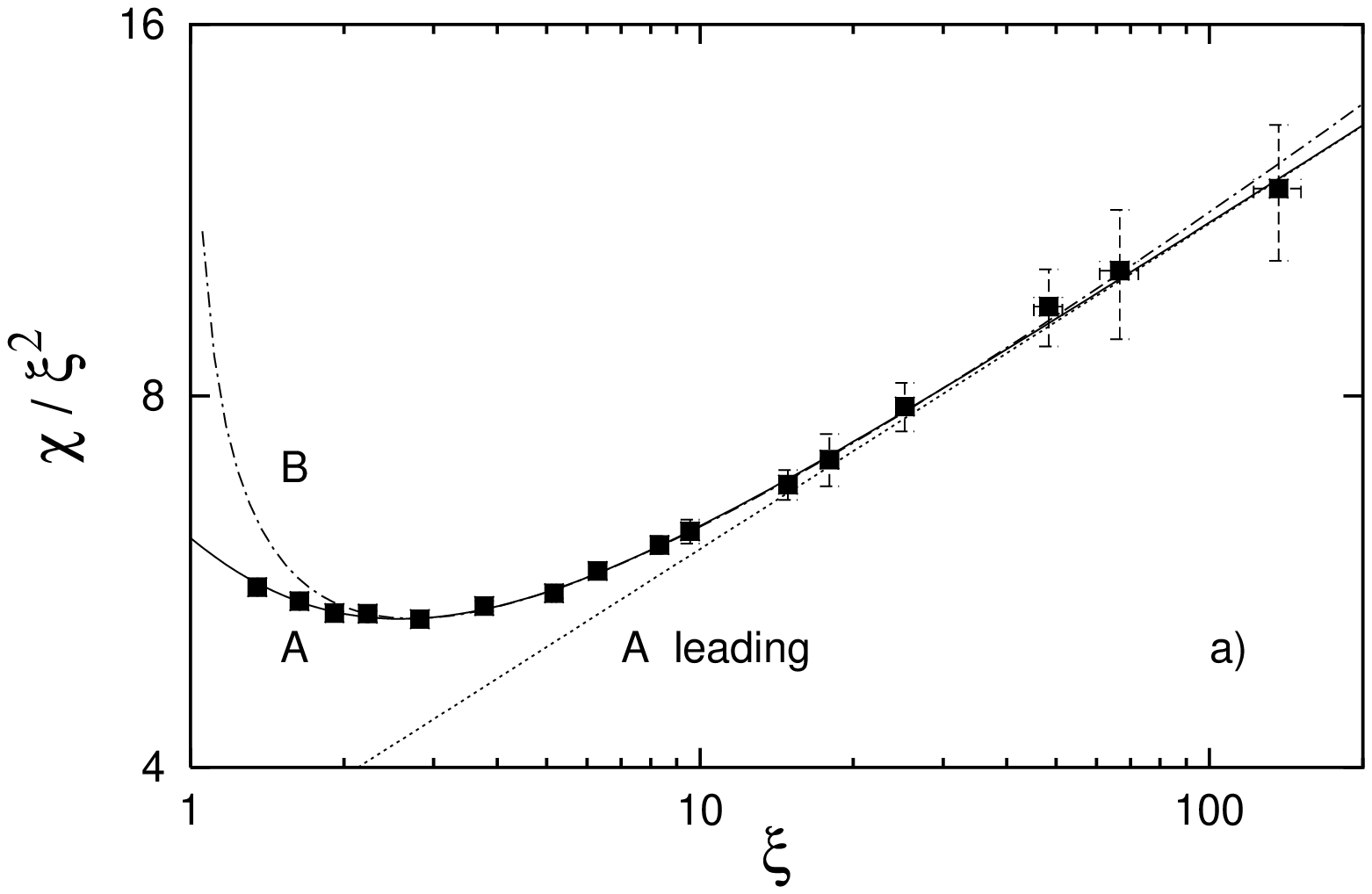}
        }
\hspace{+0.3cm}
        \mbox{
	\includegraphics[width=.32\textwidth,angle=-90]{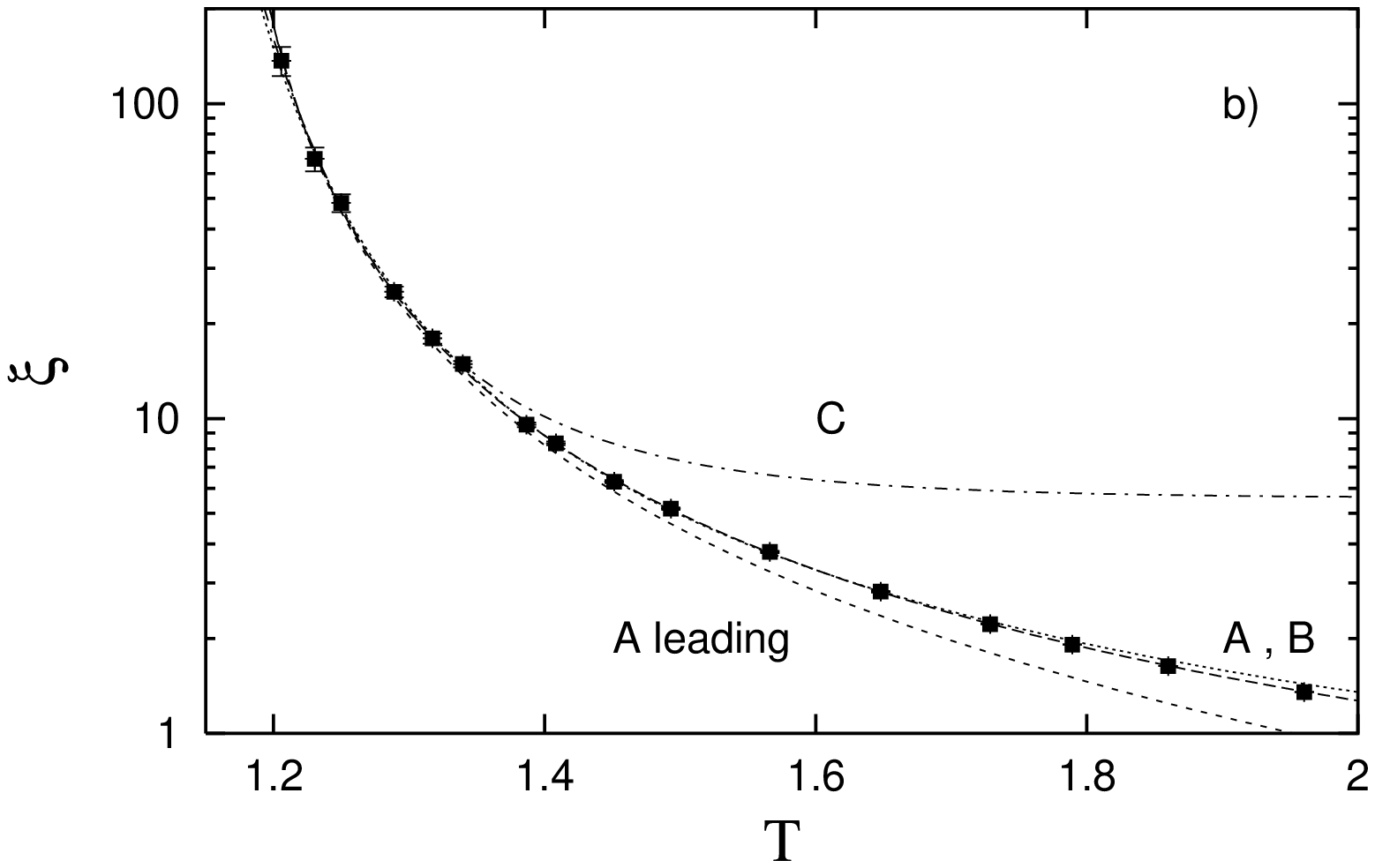}	
        }
\caption{Critical behavior of the infinite volume data. 
({\bf a}) Best fit to \protect{\reff{powerlaw2}} for $\xi\ge 1.8$ (line A),
leading term from the same fit (A leading),
best fit to \protect{\reff{logcorr}} for $\xi\ge 2.2$ (B).
({\bf b}) 
Best fit to \protect{\reff{powerlaw1}} for $\xi \ge 1.9$ (A), 
leading term from the same fit (A leading), 
best fit to \protect{\reff{KT1}} for $\xi \ge 3.8$ (B),
and best fit to (\protect{\ref{lcd}}) for $\xi \ge 14$ (C).}
\label{fitchixi}
\end{figure}

\noindent
(i) {\em  Power--law singularity at $T_c\neq 0$.}
We fit our data to 
\beqn
\xi(T) &=& c_\xi \, (T-T_c)^{-\nu} \, \left[1 + a_\xi \, (T-T_c)^\theta \right]\label{powerlaw1} \\
\chi_{SG}(\xi) &=& b \, \xi^{2-\eta} \, \left[ 1 + d \, \xi^{-\Delta} \right] 
\label{powerlaw2} 
\eeqn
with fixed $\theta$ and $\Delta$. 
Without corrections to scaling ($a_\xi=d=0$), the fit parameters 
show small {\em systematic} variations when we vary the fit interval.
The fits stabilize with $1\le\theta\le 2$ and $1\le\Delta\le 1.5$, 
the preferred values being 
$\theta=1.4$  and $\Delta=1.3$ ({\em goodness of fit} parameter 
$Q > 0.6$ and  $Q > 0.98$ resp.). 
Our estimates  for the critical parameters are $T_c = 1.156 \pm 0.015$, 
$\nu = 1.8 \pm 0.2$, and $\eta = -0.26 \pm 0.04$. 
Corrections to scaling are important for $\xi \le 10$ (Fig.\ref{fitchixi}). 

\noindent
(ii) {\em Essential singularity at $T_c\neq 0$}.
Our data fit very well also to
\beq
\xi(T) = f_\xi \, \exp \bigl( g_\xi / (T-T_c)^\sigma \bigr) \; .\label{KT1}
\eeq
For $\xi \geq 3.8$ the best fit (shown in Fig.\ref{fitchixi}b) gives 
$\sigma = 0.5 \pm 0.3$, $T_c = 1.08 \pm 0.04$ ($Q=0.69$). 
Deviations from this fit for $\xi < 3$ are 
consistent with corrections to scaling of $\approx 10\%$.
Since for an exponential singularity
we expect multiplicative logarithmic corrections,  
we tried also the fit
\beq
\chi_{SG}(\xi) = b_l \, \xi^{2-\eta_l} \, \left(\log \xi\right)^{r} \label{logcorr}
\eeq
obtaining $\eta_l = -0.36 \pm 0.03$ and
$r = -0.36 \pm 0.06$ ($Q > 0.9$) (see Fig.\ref{fitchixi}a).

\noindent
(iii) {\em Essential singularity at $T = 0$}.
Fitting our data to
\beq
\xi(T) = f_\xi \, \exp \bigl(g_\xi/T^\sigma \bigr) 
\label{lcd}
\eeq
we find $\sigma \geq 9$. Such a high value is highly implausible
on the basis of renormalization group arguments, from which we
expect $\sigma \approx 2 $  \cite{sg3d}. 
We therefore believe that an essential singularity at $T=0$ is excluded.

\vspace{-0.5cm}
\subsection*{Acknowledgements } 
We thank A.~Pelissetto and A.P.~Young for useful discussions.
This work was supported by the INFM Parallel Computing Initiative.

\clearpage
\addcontentsline{toc}{section}{Index}
\flushbottom
\printindex

\end{document}